\documentclass{report}
\usepackage{RR}
\RRNo{6326} 

\usepackage{epsfig}

\usepackage{amssymb,amsmath,amsfonts,latexsym,bbm}
\usepackage{ae,aecompl}
\usepackage{graphicx}

\newcommand{\remove}[1]{}

\newtheorem{fact}{Fact}

\usepackage[english]{babel}
\usepackage[latin1]{inputenc}
\usepackage{subfigure}
\usepackage{latexsym}

\usepackage[T1]{fontenc}

\usepackage{graphicx}
\usepackage{graphics}
\usepackage{algorithm}
\usepackage{algorithmic}

\newtheorem{lemma}{Lemma}
\newtheorem{theorem}{Theorem}
\newtheorem{corollary}{Corollary}

\floatstyle{ruled}
\newfloat{Algorithm}{thp}{lop}[section]

\newenvironment{proof}{\textbf{Proof:}}{\hfill$\Box$}

\newcounter{linecounter}

\begin{document}

\RRNo{9999}

\RRdate{Octobre 2007}

\RRtitle{Localisation déterministe et sécurisée dans les réseaux de capteurs}
\RRetitle{Deterministic Secure Positioning in Wireless Sensor Networks}
\titlehead{Deterministic Secure Positioning}

\RRauthor{Sylvie Delaët\thanks{Univ. Paris-Sud XI, France} \and Partha Sarathi Mandal\thanks{INRIA Futurs \& Univ. Paris-Sud XI, France} \and Mariusz Rokicki\thanks{CNRS \& Univ. Paris-Sud XI, France} \and Sébastien Tixeuil\thanks{Univ. Pierre \& Marie Curie, INRIA Futurs, France}}

\RRabstract{Properly locating sensor nodes is an important building block for a large subset of wireless sensor networks (WSN) applications.  
As a result, the performance of the WSN degrades significantly when misbehaving nodes report false location and distance information in order to fake their actual location.
In this paper we propose a general distributed deterministic protocol for accurate identification of faking sensors in a WSN. Our scheme does \emph{not} rely on a subset of \emph{trusted} nodes that are not allowed to misbehave and are known to every node in the network. Thus, any subset of nodes is allowed to try faking its position. As in previous approaches, our protocol is based on distance evaluation techniques developed for WSN.

On the positive side, we show that when the received signal strength (RSS) technique is used, our protocol handles at most $\lfloor \frac{n}{2} \rfloor-2$ faking sensors. Also, when the time of flight (ToF) technique is used, our protocol manages at most $\lfloor \frac{n}{2} \rfloor - 3$ misbehaving sensors.
On the negative side, we prove that no deterministic protocol can identify faking sensors if their number is $\lceil \frac{n}{2}\rceil -1$.
Thus our scheme is almost optimal with respect to the number of faking sensors.

We discuss application of our technique in the trusted sensor model. More precisely our results can be used to minimize the number of trusted sensors that are needed to defeat faking ones.
}

\RRkeyword{Wireless Sensor Network, Secure Positioning, Distributed Protocol, Faking Sensor.}

\RRresume{Localiser correctement des capteurs autonomes est une brique de base importante pour un grand nombre d'applications dans les réseaux de capteurs (WSN). En effet, l'efficacité du WSN est significativement dégradée quand des n\oe uds malicieux rapportent de fausses positions et de fausses informations de distance de manière à simuler une localisation fictive. 
Dans cet article, nous proposons une solution algorithmique distribuée pour l'identification exacte des capteurs malicieux dans un WSN. Notre approche n'est pas basée sur l'utilisation d'un sous-ensemble de n\oe uds ``de confiance'' qui serait connu de chaque autre n\oe ud du WSN. Ainsi, tout sous-ensemble des participants peut essayer de tricher sur sa position. Comme dans les approches précédentes, notre protocole est basé sur des techniques d'évaluation des distances développées pour les WSN.

Nous montrons que quand la technique de la force du signal reçu (RSS) est utilisée, notre protocole peut tolérer au plus $\lfloor \frac{n}{2} \rfloor-2$ n\oe uds malicieux. De plus, quand la technique du temps de vol (ToF) est utilisée, notre protocole peut gérer au plus $\lfloor \frac{n}{2} \rfloor - 3$ tricheurs. 
Nous montrons également qu'il est impossible pour un protocole déterministe d'identifier les n\oe uds malicieux si leur nombre est au moins égal à $\lceil \frac{n}{2}\rceil -1$, ce qui rend notre résultat presque optimal en ce qui concerne le nombre de n\oe uds malicieux tolérés.

Nous discutons l'application de notre technique au modèle où il existe des n\oe uds de confiance. Plus précisément, nos résultats peuvent être utilisés pour minimiser le nombre de n\oe uds de confiance nécessaires à la détection sans faille des n\oe uds malicieux.}

\RRmotcle{Réseaux de capteurs sans fil, localisation sécurisée, algorithme distribué, capteurs malicieux.}

\RRprojet{Grand Large}

\RRtheme{\THNum}

\URFuturs

%\begin{document}

\makeRR

\chapter{Introduction}

Properly locating sensor nodes is an important building block for a large subset of wireless sensor networks (WSN) applications.  
For example, environment and habitat monitoring~\cite{habitat}, surveillance and tracking for military~\cite{Military} or civilian purpose, both require the knowledge of the location where a particular event takes place. 
Location of nodes in a WSN can also be used for location based routing algorithms (such as geographic routing \cite{GeoRouting}), or location based services. 

Most of existing position verification protocols rely on distance evaluation techniques (\emph{e.g.} \cite{POS4, UWB, POS5, POS3, POS1, POS2}).
Received signal strength (RSS)~\cite{POS4} and time of flight (ToF)~\cite{UWB} techniques are relatively easy to implement yet very precise (one or two meters).
In the RSS technique, receiving sensor estimates the distance of the sender on the basis of sending and receiving signal strengths.
In the ToF technique, sensor estimates distance based on message delay and radio signal propagation time.
Position verification using the aforementioned distance estimation techniques is relatively straighforward provided that \emph{all} sensors cooperate.
However, this task becomes challenging in the presence of misbehaving nodes that are allowed to report false position and distance information in order to fake their actual position. In the following such nodes are denoted as \emph{faking} or \emph{cheating} nodes. 

Such misbehaviors could occur due to several factors: a sensor may malfunction due to improper sensor deployment, partial communication problem due objects in the vicinity, or inaccurate position (coordinates) estimation. We consider that misbehaving sensors are unaware that they are malfunctioning, so locally they properly execute the protocol that is given to all nodes. Nevertheless, they can report incorrect position, change signal strength (when the RSS technique is used), or report incorrect transmission time (when the ToF technique is used).

\section{Related Work}
Most methods~\cite{CCS06,CH06,LCP06,LP04} existing in the literature that use distance estimation techniques to detect and filter out faking nodes are based on the availability of a few fixed trusted entities (or \emph{verifiers}), that are equipped with GPS. We refer to this model as the \emph{trusted sensor} (or \emph{TS}) model. 
In this model, the faking nodes may use attacks not available to regular nodes, such as radio signal jamming or using directional antenas, that permit to implement \emph{e.g.} wormhole attack \cite{HPJ03} and Sybil attack \cite{D02}.
Lazos and Poovendran~\cite{LP04} present a secure range-independent localization scheme, where each sensor computes its position based on received beacons messages from locators. Sensors compute the center of gravity of beacons's intersection region, and the computed location becomes the estimated location of the sensor. Probabilistic analysis of the protocol demonstrate that it is resilient to wormhole and Sybil attacks, with high probability. Lazos \emph{et al.}~\cite{LCP06} further refine this scheme with multilateration to reduce the number of required locator, while maintaining probabilistic guarantees.
The protocol of Capkun and Hubaux \cite{CH06} relies on a distance bounding technique proposed by Brands and Chaum \cite{DistBound}.
Each sensor $v$ measures its distance to a (potential) faking sensor $u$ based on its message round-trip delay and radio signal propagation time, thus enabling the faking node $u$ only to \emph{enlarge} the distance to $v$. Then, if the faking node is located inside the triangle formed by verifiers and its faked position is also
located within the triangle, then at least one of the three verifiers detects an inconsistency.
Capkun, Cagalj, Srivastava \cite{CCS06} is supported by powerful verifiers, that know their positions and communicate with some wired channels that prevent faking nodes to locate them or to listen their transmissions. Then, each verifier $v$ measures the arrival time $t_v$ of the (potential) faking node transmission. Verifiers exchange all such arrival times and check consistency of the declared position. 
 However, the TS model presents several drawback in WSNs: first the network can not self-organize in an entirely distributed manner, and second the trusted nodes have to be checked regularly and manually to actually remain trusted. 

Relaxing the assumption of trusted nodes makes the problem more challenging, and to our knowledge, has only been investigated very recently~\cite{HHK07}. We call this model where no trusted node preexists the \emph{no trusted sensor} (or \emph{NTS}) model.
The approach of~\cite{HHK07} is randomized and consists of two phases: distance measurement and filtering.
In the distance measurement phase, sensors measure their distances to their neighbors, faking sensors being allowed to corrupt the distance measure technique.
In the filtering phase each correct sensor randomly picks up $2$ so-called \emph{pivot} sensors.
Next each sensor $v$ uses trilateration with respect to the chosen pivot sensors to compute the location of its neighbor $u$. If there is a match between the announced location and the computed location, the $(u,v)$ link is added to the network, otherwise it is discarded.
Of course, the chosen pivot sensors could be faking and lying, so the protocol may only give probabilistic guarantee.

In this paper we present a deterministic protocol that performs in the NTS model and where every correct (\emph{i.e.} non faking) node: \emph{(i)} identifies the positions (coordinates) of all correct nodes, and \emph{(ii)} identifies the faking nodes (if any). The goal of the faking nodes is to convince the correct nodes that they are located in a fake position.

\section{Our results} 
The main contribution of this paper is a secure deterministic positioning protocol, \textsc{FindMap}, in the NTS model. 
To the best of our knowledge, it is the first deterministic protocol for this problem in the NTS model.
The basic version of the protocol assumes that faking sensors are not able to mislead distance evaluation techniques.
Then, our protocol correctly filters out faking sensors provided they are at most $\lceil \frac{n}{2} \rceil -2$.
Conversely, we show evidence that it in the same setting, it is impossible to deterministically solve the problem when the number of faking sensors is at least $\lceil \frac{n}{2} \rceil -1$.
We then extend the protocol do deal with faking sensors that are also allowed to corrupt the distance measure technique (RSS or ToF). 
In the case of RSS, our protocol tolerates at most $\lfloor \frac{n}{2} \rfloor-2$ faking sensors (provided that no four sensors are located on the same circle and no three sensors are co-linear).
In the case of ToF, our protocol may handle up to $\lfloor \frac{n}{2} \rfloor - 3$ faking sensors (provided that no six sensors are located on the same hyperbola and no three sensors are co-linear).

Our results have significant impact on secure positioning in the TS model as well.
The TS protocol presented by Capkun \emph{et al.}~\cite{CCS06} relies on set of hidden stations, that detect inconsistencies between measured distance and distance computed from claimed coordinates, using ToF-like technique to estimate the distance. Our detailed analysis shows that six hidden stations (verifiers) are sufficient to detect inconsistency in the same setting. In~\cite{CCS06}, the authors conjecture that the ToF-like technique could be replaced with RSS technique. Our results anwser positively to the open question of~\cite{CCS06}, improving the number of needed stations to four. So, in the TS model, our results can be used to efficiently deploy a minimal number trusted stations.

\chapter{Technical preliminaries}
\label{sec:tech}

We assume that every node is able to communicate to every other node in the WSN. 
The size of the WSN is $n$ and is known to every node. 
Each node is also aware of its own geographic coordinates, and those coordinates are used to identify nodes.
The WSN is partially synchronous: every node operates in rounds. 
In one round, every node is able to send exactly one message to every other node wihout collision occuring.
For each transmission, a correct nodes uses the same transmission power $S_s$.
 
Faking nodes are allowed to transmit incorrect coordinates (and thus incorrect identifier) to the other nodes.
In the basic protocol, faking nodes can not corrupt distance measure techniques, while in Section \ref{sec:RSS} we relax this assumption and allow faking sensors to change its radio transmitter power and send a related fake position to the correct nodes.
In Section \ref{sec:ToF} a faking sensor also can report incorrect transmission time. 
Also, we assume that faking nodes may cooperate between themselves in an omniscient manner (\emph{i.e.} without exchanging messages) in order to fool the correct nodes in the WSN.

We assume that all distance estimation techniques are perfect with respect to precision.
The distance computed by node $v$ to node $u$ based on a distance estimation technique is denoted by $\hat{d}(v,u)$.
The distance computed by $v$ to the node $u$ using coordinates provided by $u$ is denoted by $d(v,u)$.
A particular sensor $v$ \textit{detects inconsistency} on distance (\emph{i.e.} position) of sensor $u$ if  $d(v,u) \neq \hat{d}(v,u)$.
Our protocols rely on detecting and reporting such inconsistencies.

In the remaining of the paper, we use three distance estimation techniques:
\begin{enumerate}

\item In the \emph{received signal strength} (\emph{RSS}) technique we assume that each node can precisely measure the distance to the transmitting node from RSS by Frii's transmission equation \ref{FriiEq}~\cite{LF88}:
\begin{eqnarray}
\label{FriiEq}
S_r = S_s \left(\frac{\lambda}{4 \pi d}\right)^2
\end{eqnarray}
Where $S_s$ is the transmission power of the sender, $S_r$ is the remaining power or receive signal strength (RSS) of the wave at receiver, $\lambda$ is wave length and $d$ is distance between sender and receiver.

\item The \emph{synchronous time of flight} (\emph{SToF}) technique relies on propagation time of the radio signal.
For this technique we assume that sensors are synchronized by global time. Sender $u$ attaches the time of transmission, $t_s$ to the message. The receiver $v$ records the message arrival time $t_r$ of the message. Next $v$ computes the distance $d = t*s$ of $u$ based on time delay $t=t_r-t_s$ of the message and radio signal speed $s$. 

\item The \emph{different arrival time} (\emph{DAT}) technique provides similar guarantees as SToF.
The advantage of DAT over SToF is that DAT does not require synchronization. In the DAT technique each sensor transmits its message with two types of signals
that differ on propagation speed \emph{e.g.} radio signal (RF) and ultra sound signal (US).
Sender sensor $u$ transmits its message with RF and US signal simultaneously.
Receiver sensor $v$, which estimates its distance to sender $u$,
records arrival time $t_r$ of RF signal and arrival time $t_u$ of US signal from $u$.
Then, based on the propagation speed $s_r$ of RF, propagation speed $s_u$ of US and difference of arrival times $t=t_u-t_r$ sensor $v$ can compute distance to sensor $u$. Equation \ref{DATeq} show the relation.
\begin{eqnarray}
\label{DATeq}
t = \frac{\hat{d}}{s_r} - \frac{\hat{d}}{s_u}
\end{eqnarray}

\end{enumerate}

\chapter{Basic Protocol}
\label{sec:Protocol}

In this section we present the protocol \textsc{FindMap}, that essentially performs by majority voting.
The protocol detects all faking sensors provided that $n-2-f>f$.
Thus the total number of faking sensors is at most $\lceil \frac{n}{2} \rceil -2$.
In this section we consider the relatively simpler case where faking sensors are not able to cheat the distance estimation techniques (see above) that are used by the correct nodes. 
Our second key assumption is that no three correct sensors are co-linear.
This assumption allows to formulate the following fact.

\begin{figure}
\begin{center}
\epsfig{figure=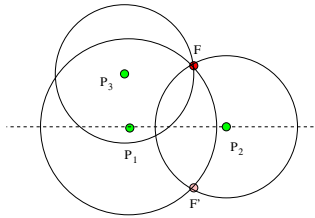 , width=0.3\textwidth}
\caption{\label{fig:fig0} Example in which sensor $F$ consistently fakes its location to $F'$ against sensors $P_1$ and $P_2$. However the third sensor $P_3$ always detects an inconsistency since no three correct sensors are co-linear.}
\end{center}
\end{figure}

\begin{fact}
\label{fact:3DetectInconsitency}
If a faking sensor transmits a message with a fake position then at least one of three correct sensors can detect an inconsistency (see Figure \ref{fig:fig0}).
\end{fact}

Based on Fact \ref{fact:3DetectInconsitency}, we can develop \textsc{FindMap(\textit{threshold})}.
The protocol operates in two rounds. The protocol is paremeterized by a \textit{threshold} parameter.
In \textit{Round 1} all sensors exchange their  coordinates by transmitting an initial message.
Next each node $v$ computes the distances $\hat{d}(v,u)$ (from the distance estimation technique) and $d(v,u)$ (from the obtained node coordinates) of $u$ and compare them.
If $\hat{d}(v,u) \neq d(v,u)$ then $v$ accuses $u$ to fake its position. Otherwise $v$ does not accuse $u$. To keep record of its accusations, each node $v$ maintain an array $accus_v$ of size $n$. In \textit{Round 2} each node $v$ exchanges its array of accusations. 
Next each node $v$ counts accusations toward every other node $u$ including its own accusations. A sensor $v$ detects a sensor $u$ as faking if the number of accusations is at least equal to the threshold parameter. For our basic \textsc{FindMap} protocol we use $\textit{threshold}=\lfloor \frac{n}{2} \rfloor$.

{\underline{\bf Protocol FindMap($\textit{threshold} = \lfloor \frac{n}{2} \rfloor$)}}\\
\framebox[15cm][t]{\parbox[t][][t]{16.5cm}{\vspace{-.5cm}
\begin{tabbing}
\=xxx\=xxx\=xxx\=xxx\=xxx\=xxx\=xxxx\=xxxx\=xxxx\=xx\=xx\=\kill
{\it Round 1:}\\
1. \>\>$v$ exchange coordinates by transmiting $init_v$ and receiving $n-1$ messages.\\
2. \>\>for each received message $init_u$:\\
3. \>\>\>compute $\hat{d}(v,u)$ and $d(v,u)$ using the coordinates of $u$.\\
4. \>\>\>   {\bf if} ($\hat{d}(v,u)\neq d(v,u)$) {\bf then} $accus_v[u] \leftarrow true$ \\
5. \>\>\>  {\bf else} $accus_v[u] \leftarrow false $\\
{\it Round 2:}\\
6. \>\>$v$ exchange accusations by transmiting $accus_v$ and receiving $n-1$ accusations. \\
7. \>\>for each received $accus_u$: \\
8. \>\>\>for $r = 1~ to~ n$ \\
9. \>\>\>\> {\bf if} $accus_u[r] = true$ {\bf then}   $NumAccus_r +=1$ \\
10. \>\>for each sensor $u$: \\
11. \>\>\> {\bf if} ($\textit{threshold} \le NumAccus_u$) {\bf then} $v$ considers $u$ is faking.
\end{tabbing}\vspace{-.5cm}
}}

\begin{theorem}
\label{Thm_PosRes}
Protocol \textsc{FindMap($\lfloor \frac{n}{2} \rfloor$)} identifies all the faking sensors and finds the position of correct sensors provided $n-f-2 > f$.
\end{theorem}

\begin{proof}
First we will show that each faking sensors will be accused by proper number of correct sensors.
In each subset of three correct sensors there exists at least one which detects inconsistency
on distance to a faking sensors.
This is guaranteed by fact \ref{fact:3DetectInconsitency}.
Thus each faking sensors will be accused by at least $n-f-2$ correct sensors.
Inequality $n-f-2 > f$ guarantees that number of correct sensors is at least $\lfloor \frac{n}{2} \rfloor$.
We can also observe that each correct sensors can be accused by at most
$\lceil \frac{n}{2} \rceil -2$ faking sensors.
However this is not enough to find a correct sensors faking.
\end{proof}

Next we show that it is impossible to detect the real location of correct sensors and filter out the faking one when $n-2-f \leq f$.
The assumption that faking sensors cannot corrupt the distance ranging technique makes this result even stronger. Our protocol is synchronous but this impossibility result holds for asynchronous settings too.

\begin{theorem}
\label{Thm_ImpRes}
If  $n-f-2 \leq f$ then the real location of the correct sensors cannot be detected by a deterministic protocol.
\end{theorem}

\begin{proof}
\begin{figure}[htbp]
\begin{center}
\epsfig{figure=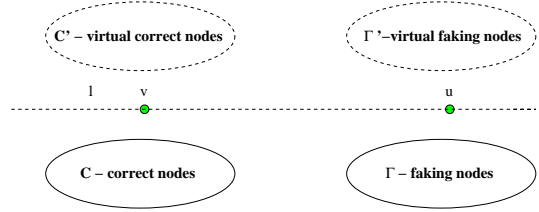 , angle=270, width=0.5\textwidth}
\caption{\label{fig:fig1} First execution.}
\end{center}
\end{figure}
Let us assume that correct sensors run a protocol $\mathcal{P}$,
which allows to detect location of correct sensors
and identify the faking sensors even when $n-f-2=f$.
In case $n-f-2<f$ we make some faking sensors correct to achieve equality
and in case $n$ is odd one of the faking sensors will remain silent.
Let us consider the first execution (see figure \ref{fig:fig1}).
There are two correct sensors $v$ and $u$ located on the straight line $l$.
There are two sets of sensors $C$-correct sensors and $\Gamma$-faking sensors located on the lower half of the plane.
The sizes of the sets are equal $|C|=|\Gamma|=f$.
The sensors in $\Gamma$ are trying to convince sensors $v$ and $u$ that
they are located in $\Gamma'$ on the other side of the straight line $l$ symmetrically.
Each sensor in $\Gamma$ behave as if it was a correct sensor reflected symmetrically against straight line $l$. 
The sensors in $\Gamma'$ are called virtual faking sensors.
Virtual sensors in $\Gamma'$ execute the protocol as if sensors in $C$ were faking and their correct location was in $C'$,
which is symmetric reflection of $C$ against straight line $l$.
Construction of the second execution will clarify why we need such behavior of sensors in $\Gamma'$.
We can see that sensors $v$ and $u$ are not able to detect inconsistency directly on the distance
of virtual faking sensors since symmetry preserves their distances from $v$ and $u$.
By our assumption about correctness of the protocol $\mathcal{P}$ sensors $v$ and $u$ are able
to verify that sensors in $\Gamma'$ are faking.

\begin{figure}[htbp]
\begin{center}
\epsfig{figure=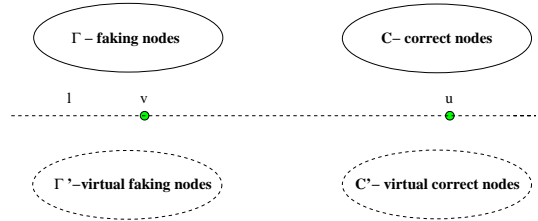 , angle=270, width=0.5\textwidth} \caption{Second execution.}
\label{fig:fig2}
\end{center}
\end{figure}

Now let us consider the second execution (see figure \ref{fig:fig2}).
In the second execution sensors in $C$ and $\Gamma'$ are swapped.
Thus sensors in $\Gamma$ has to be located on the other side of straight line $l$  symmetrically.
Now virtual faking sensors in $\Gamma'$ can imitate the first execution of the correct sensors in $C$.
Correct sensors in $C$ behave like virtual sensors in $\Gamma'$ in first execution.
This is because the virtual sensors in $\Gamma'$ in the first execution behaved like correct sensors
and additionally they claimed that sensors from $C$ were located in $C'$ (see figure \ref{fig:fig2}).
Now $\Gamma$ is really located in the previous location of $C'$ and the sensors in $C$ are correct.
Thus sensors $v$ and $u$ are not able to distinguish between the first and the second execution.
Sensors $v$ and $u$ will have to decide that $C$ is set of faking sensors.
This is because $v$ and $u$ have made such decision in first execution and $v$ and $u$ is not able
to distinguish between these two executions.
\end{proof}

\chapter{Protocol based on RSS ranging technique}
\label{sec:RSS}

In this section, we consider that sensors use RSS technique to measure distance.
We are assuming that each correct sensor has a fixed common transmission signal strength of $S_s$.
The faking sensors can change their transmission signal strength and send suitable fake position to other sensors.  
Let $F$ be a faking sensor that changes its signal strength $S^{'}_{s}$ and sends a suitable fake position $F'$ to other correct sensors. Sensor $v$ can estimate the distance, $\hat{d}$ from the receive signal strength (RSS) by Frii's transmission equation assuming the common signal strength $S_s$ has been used, according to the assumption in section \ref{sec:tech}.
$${\hat{d}}^2= c \frac{ S_s}{S_r} \implies {\hat{d}}^2 = \frac{S_s}{S^{'}_s} d^2\ldots (2)$$
where $c=\left(\frac{\lambda }{ 4 \pi}\right)^2$, $S_r = c \frac{ S^{'}_{s}}{d^2}$, and $d$ is the distance from $v$ to the actual position of $F$.

We show that Protocol \textsc{FindMap($\lceil \frac{n}{2}\rceil -1$)} can be adapted to this model provided that $n-3-f > f$, \emph{i.e.} the total number of faking sensors is at most $\lfloor \frac{n}{2} \rfloor -2$ and no four correct sensors are located on a particular circle. In this variant of the protocol, a sensor $v$ considers sensor $u$ faking if the number of accusations messages for $u$ is at least $\lceil \frac{n}{2} \rceil -1$.

\begin{lemma}
\label{L1}
Let $F$ be a faking sensor, and $P_1$ and $P_2$ be two correct sensors. There exists a position $(x_f, y_f)$ for $F$ such that $F$ is always able to fake a position $F'=(x'_f,y'_f)$ to both $P_1$ and $P_2$, with $x_f\neq x'_f$, and $y_f\neq y'_f$ by changing its signal strength from S to $S'$.
\end{lemma}
\begin{proof}
\begin{figure}[htbp]
\begin{center}
\epsfig{figure=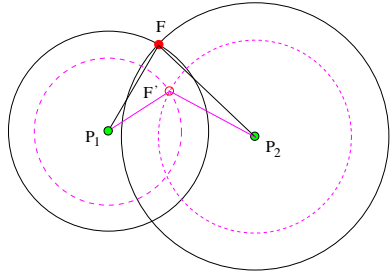,width=2in} \caption{An example showing a
faking sensor $F$ can supply its suitable false position $F'$ to
correct sensors $P_1$ and $P_2$ by changing its signal strength.}
\label{fig1}
\end{center}
\end{figure}
The faking sensor, $F$ changes its signal strength from $S_s$ to
$S^{'}_{s}$ and sends a corresponding fake position $(x'_f,y'_f)$ to
$P_1$ and $P_2$ such that $${\hat{d_1}}^2 = \frac{S_s}{S^{'}_s}
{d_1}^2~~ {\rm and}~~ {\hat{d_2}}^2 = \frac{S_s}{S^{'}_s} {d_2}^2$$
Where $\hat{d_1}$ and $\hat{d_2}$ are the estimated distances
measured by $P_1$ and $P_2$ respectively from the RSS of $F$ and
$(x'_f,y'_f)$ is the point of intersection of the two circles
centering at $P_1$ and $P_2$ with radius $\hat{d_1}$ and $\hat{d_2}$
respectively according to the figure \ref{fig1}, $d_1$ and $d_2$ are
the distances from the actual position $(x_f,y_f)$ of $F$ to $P_1$
and $P_2$ respectively

Then $P_1$ and $P_2$ can not able to detect the inconsistency of the
fake position $(x'_f,y'_f)$ of $F$ such that $x_f\neq x'_f$, and
$y_f\neq y'_f$.\end{proof}

\begin{lemma}
\label{L2} Let $F$ be a faking sensor, and $P_1$ and $P_2$ be two
correct sensors. There exists a position $(x_f,y_f)$ for $F$ such that
$F$ can always choose a fake position $F' =(x'_f,y'_f)$ for both $P_1$
and $P_2$, with $x_f\neq x'_f$, and $y_f\neq y'_f$ by changing its
signal strength. Then the possible fake locations for $F'$ are placed on a circular arc.
\end{lemma}

\begin{proof}
From lemma \ref{L1} we know that  $\frac{\hat{d_1}}{d_1} =
\surd\left(\frac{S_s}{S^{'}_s}\right)~~ {\rm and}~~
\frac{\hat{d_2}}{d_2} = \surd\left(\frac{S_s}{S^{'}_s}\right)$ that
is $\frac{\hat{d_1}}{d_1}=\frac{\hat{d_2}}{d_2}$  or
$\frac{\hat{d_1}}{\hat{d_2}}=\frac{d_1}{d_2}$ implies
$\frac{\hat{d_1}}{\hat{d_2}}=\delta$ where $\delta=\frac{d_1}{d_2}$
= constant, for a pair of sensors $P_1$ and $P_2$.

\begin{figure}[htbp]
\begin{center}
\epsfig{figure=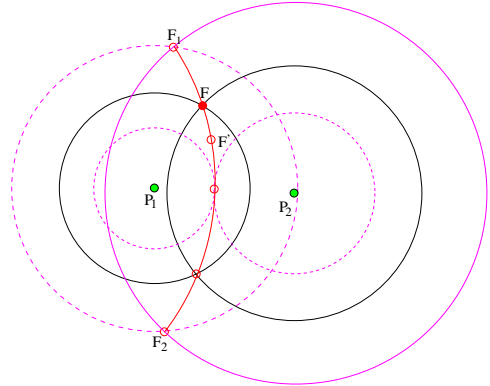,width=2.5in} \caption{An example showing possible locations ($F_1 F F_2$) of the fake position $(x'_f,y'_f)$ than can be supplied by faking sensor $F$ for a pair correct sensors $P_1$ and $P_2$ by changing its signal strength.} \label{fig2}
\end{center}
\end{figure}

If  $(x_1,y_1)$ and $(x_2,y_2)$ are the coordinates of $P_1$ and
$P_2$ then the possible location of the $(x'_f,y'_f)$ is $$\frac{(x - x_1)^2 +
(y - y_1)^2}{(x - x_2)^2 + (y - y_2)^2} = \delta^2$$
$\implies x^2 + y^2 -2
\left(\frac{x_1-\delta^2x_2}{1-\delta^2}\right)x -2\left(
\frac{y_1-\delta^2y_2}{1-\delta^2}\right)y + \frac{{x_1}^2 +
{y_1}^2- \delta^2({x_2}^2 +{y_2}^2)}{1-\delta^2}=0$

Which is an equation of circle, where $\delta = \surd\frac{(x_f -
x_1)^2 + (y_f - y_1)^2}{(x_f - x_2)^2 + (y_f - y_2)^2}$.

Now we have to prove that $(x'_f,y'_f)$ can lay only on $F_1 F F_2$
part of circular arc as shown in figure \ref{fig2}. Where $F_1$ and
$F_2$ are the point of intersection of two circle of transmission
range centering at $P_1$ and $P_2$ such that at least one of the
circles is its maximum transmission range.

We can prove this by contradiction. Suppose, $(x'_f,y'_f)$ laying on
the counterpart of the circular arc $F_1 F F_2$. Then it is not
possible by $F$ to pretend its fake position to $P_1$ and $P_2$
simultaneously. Since counterpart of the circular arc $F_1 F F_2$
does not belong to the common transmission of $P_1$ and $P_2$, hence
proved.\end{proof}

\begin{lemma}
\label{L3} Let $F$ be a faking sensor, and $P_1$, $P_2$, $P_3$ be
three correct sensors on a circle. There exists a position $(x_f,y_f)$
for $F$ and positions $(x_1,y_1)$, $(x_2,y_2)$  and $(x_3,y_3)$ such
that $F$ is always able to fake a position $F'=(x'_f,y'_f)$ to $P_1$,
$P_2$ and $P_3$ such that $x_f\neq x'_f$, and $y_f\neq y'_f$.
\end{lemma}
\begin{proof}
\begin{figure}[htbp]
\begin{center}
\epsfig{figure=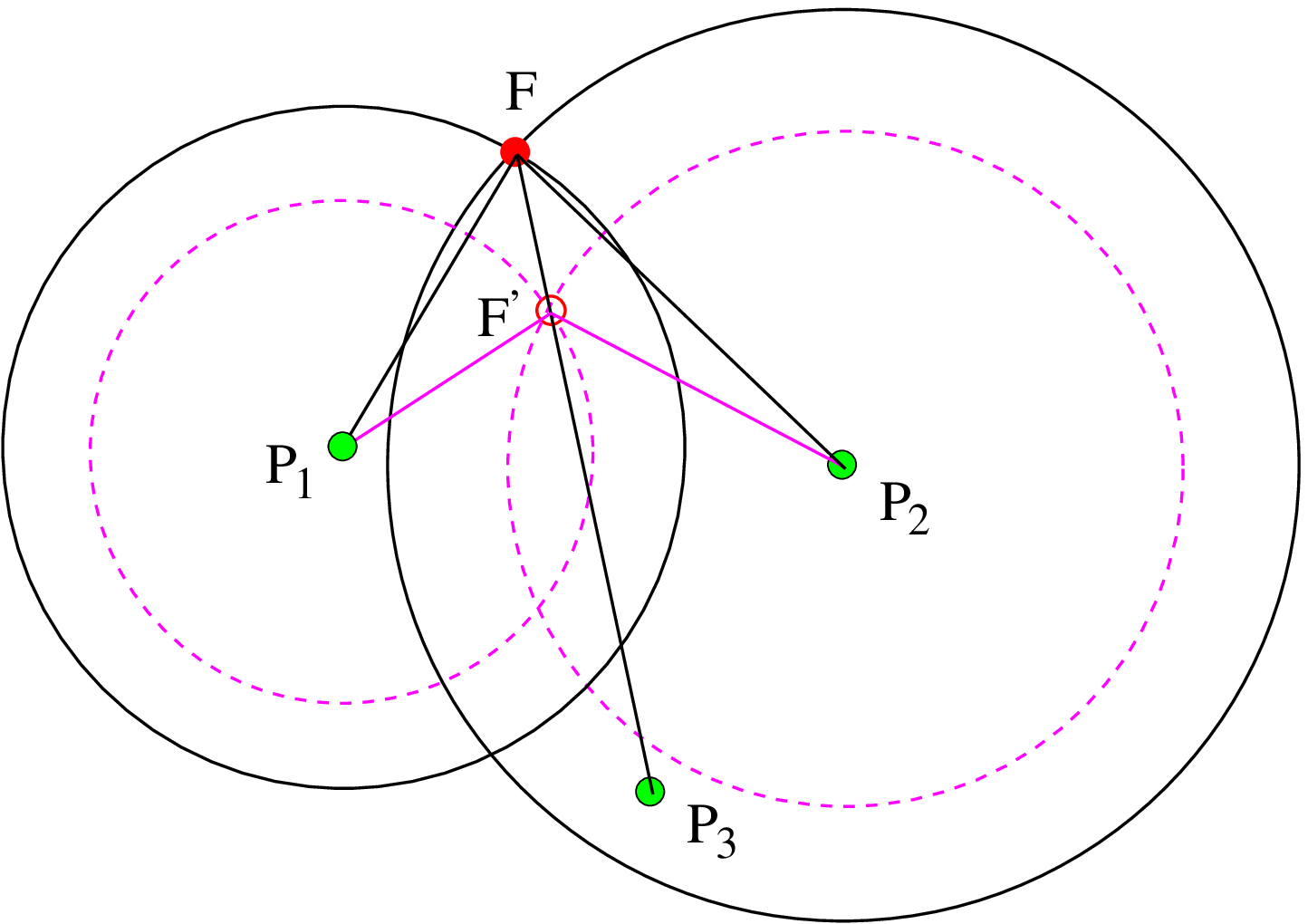,width=2.0in}
\epsfig{figure=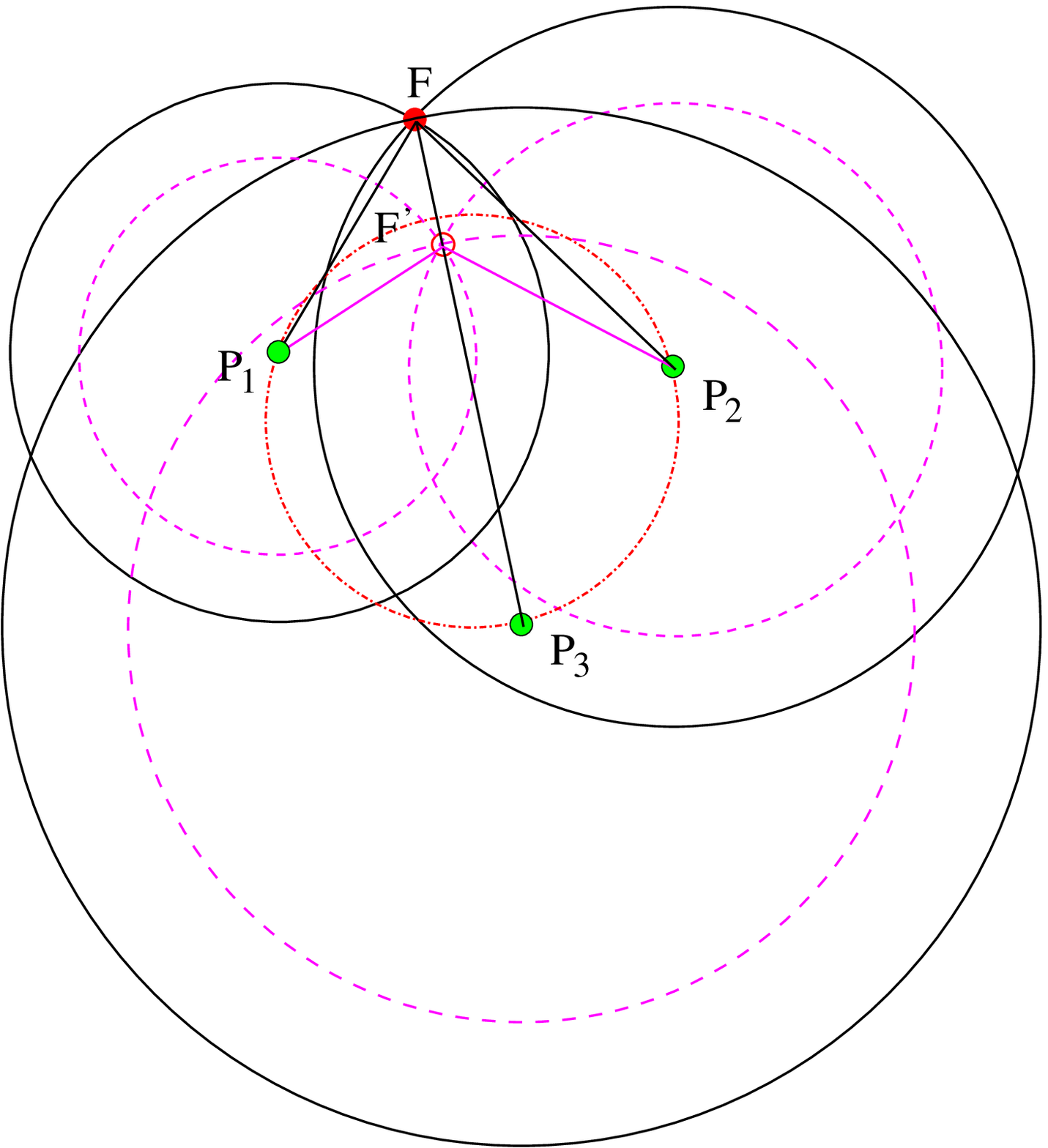,width=1.5in} \caption{An example showing a
faking sensor $F$ can lie about its position by changing signal
strength to three correct sensors.} \label{fig34}
\end{center}
\end{figure}
From Lemma \ref{L1} and \ref{L2}, faking sensor $F=(x_f,y_f)$
can fake its position $F'=(x'_f,y'_f)$ to two correct sensors $P_1$,
$P_2$ by changing its signal strength from $S_s$ to $S'_s$ such that
$P_1F':P_1F = \lambda$ and $P_2F':P_2F = \lambda$ where $\lambda
=\surd\frac{S_s}{S^{'}_s}$ and $P_1F'=\hat{d_1}$, $P_1F=d_1$,
$P_2F'=\hat{d_2}$, $P_2F=d_2$.

We have to prove that there exist a sensor $P_3$ with coordinates
($x_3,y_3$) such that $P_3$ can not able to detect the inconsistency
of fake position $(x'_f,y'_f)$, i.e., $P_3$ has to locate at a
position like $P_1$ and $P_2$ such that $P_3F':P_3F = \lambda$ as
shown in figure \ref{fig34}. Therefore $FF' : F'P_3 = (1-\lambda):
\lambda$ Therefore $(x_3, y_3) = \left( \frac{x'_f - \lambda
x_f}{1-\lambda}, \frac{y'_f - \lambda y_f}{1-\lambda}\right)$. From
geometry we know that only one circle pass through three fix points,
hence proved.\end{proof}

\begin{lemma}
\label{L4} Let $F$ be a faking sensor, and $P_1$, $P_2$ be correct
sensors. There exists a position $(x_f,y_f)$ for $F$ and positions
$(x_1,y_1)$, $(x_2,y_2)$ such that $F$ is always able to fake a
position $F'=(x'_f,y'_f)$ to $P_1$ and $P_2$ such that $x_f\neq x'_f$,
and $y_f\neq y'_f$. Then $F$ also can fake the position
$(x'_f,y'_f)$ to more $P_i$'s if and only if they lay on a particular
circle.
\end{lemma}

\begin{proof}
Lemma \ref{L2} implies that faking sensor $F$ can fix a fake
position $F'$ on the circular arc $F_1 F F_2$ with a
suitable changed signal strength ($S'$) such that $P_1$ and $P_2$ can not
able to detect the inconsistency as shown in figure \ref{fig5}.
\begin{figure}[htbp]
\begin{center}
\epsfig{figure=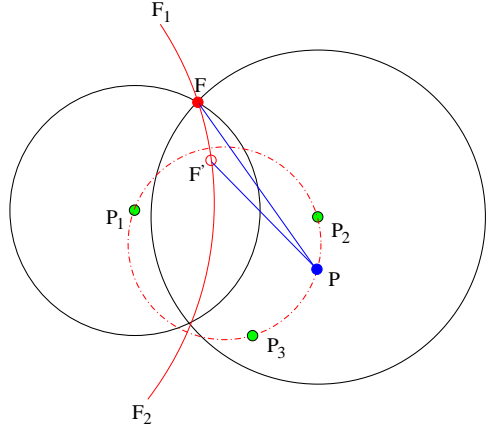,width=2.5in} \caption{An example showing a
faking sensor $F$ can lie about its position by changing signal
strength to multiple number of correct sensors which are laying on a
particular circle.} \label{fig5}
\end{center}
\end{figure}

Let $P$ is a variable point such that it keeps the same ratio
$\surd\frac{S_s}{S^{'}_s}~(= \lambda$) like $P_1$ and $P_2$ with $F$
and $F'$. Then $P$ also can not able to detect the
inconsistency of the fake position $F'$. If $\hat{d_p}$ is
the distance between $P$ and $F'$ and $d_p$ is the distance
between $P$ and $F$ then
$\frac{\hat{d_p}}{d_p} = \lambda$

Therefore the possible location of the point $P$ is $\frac{(x - x'_f)^2 + (y -
y'_f)^2}{(x - x_f)^2 + (y - y_f)^2} = \lambda^2$

$\implies x^2 + y^2 -2
\left(\frac{x'_f-\lambda^2x_f}{1-\lambda^2}\right)x -2\left(
\frac{y'_f-\lambda^2y_f}{1-\lambda^2}\right)y + \frac{{x'_f}^2 +
{y'_f}^2- \lambda^2({x_f}^2 +{y_f}^2)}{1-\lambda^2}=0$

This is an equation of circle with respect to the given fake
position $F'$ of $F$ and $P_1$ and $P_2$ as shown in
figure \ref{fig5}. Therefore, $F$ pretends the fake position
$F'$ to the sensors which are laying only on the particular circle.
\end{proof}

\begin{theorem}
\label{Thm1} Let $F$ be a faking sensor, and $P_1$, $P_2$, $P_3$ be
three correct sensors on a circle. If there exist a sensor $P_4$ which does not
lay on the same circle, $P_4$ is able to detect the inconsistency of $F$.
\end{theorem}

\begin{proof} From lemma \ref{L3} faking sensor $F$ can convey the fake position $F'$
to $P_1$, $P_2$, $P_3$, provided circles with radius $\hat{d_1} = \lambda d_1$,
$\hat{d_2} = \lambda d_2$, and $\hat{d_3} = \lambda d_3$  centering at $P_1$, $P_1$, and $P_1$ respectively
intersect at $F'$, where $\lambda =\surd\left(\frac{S_s}{S^{'}_s}\right)$.
\begin{figure}[htbp]
\begin{center}
\epsfig{figure=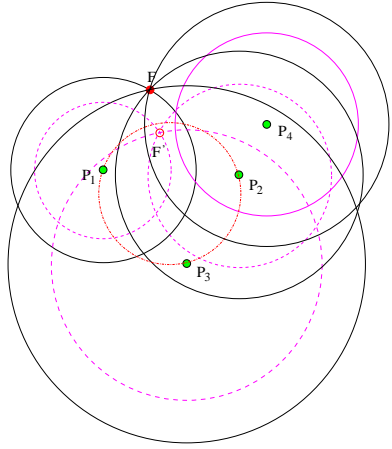,width=2in} \caption{An example showing a
that if four sensors $P_1$, $P_2$, $P_3$, $P_4$ do not lay in a particular circle then
faking sensor $F$ can  be detected by sensor $P_4$ which is not laying on the circle.} \label{fig6}
\end{center}
\end{figure}

As $P_4$ not on the circle then $\hat{d_4} \ne \lambda d_4$ as in figure \ref{fig6} implies $\hat{d_4} \ne d(P_4,F')$,
where $d(P_4,F')$ is the distance from $P_4$ to $F'$ calculated from coordinates of $F'$. Hence $P_4$ can able to detect the inconsistency of faking node $F$.
\end{proof}

\begin{corollary}
\label{Cor_RSS_FindMap}
The protocol \textsc{FindMap($\lceil \frac{n}{2} \rceil -1$)} identifies all faking sensors in the model where faking sensors can corrupt RSS ranging technique, provided that $n-f-3 > f$ and no four sensors are located on the same circle and no three sensors are co-linear.
\end{corollary}

\begin{proof}
Let us consider a faking sensor $F$, which fakes its transmission power.
Theorem \ref{Thm1} guarantees that in each set of four correct
sensors there exists a sensor, which detects inconsistency on distance to $F$.
Thus each faking sensor will be accused by at least $n-f-3$ correct sensors.
By inequality $n-f-3 > f$ the number of correct sensors that accuse $F$ is at least $\lceil \frac{n}{2}
\rceil -1$ and the number of faking sensors is at most $\lfloor \frac{n}{2} \rfloor - 2$.
Thus each faking sensor will be found faking and no correct sensor will be found faking.
If faking node $F$ does not change its transmission power but only lies about its position 
then at least one on three no-linear correct sensors will detect inconsistency.
\end{proof}

Theorem \ref{Thm1} can be also applied in the protocol for the model of trusted sensors.
In the protocol presented in \cite{CCS06}, we can use theorem \ref{Thm1} to find deployment of the minimum number of hidden stations required to detect faking nodes.

\begin{corollary}
\label{hiddenBase_RSS}
If the four hidden stations are not located on the same circle and no three stations are co-linear then one of the stations will always detect a faking node.
\end{corollary}

Corollary~\ref{hiddenBase_RSS} remains true provided the faking node's transmission reaches all hidden stations and it is not allowed to use directional antennas.
Since the verifiers are hidden to the faking node in the model of \cite{CCS06}, the latter has very low chances to consistently fake its position even with directional antennas.

\chapter{Protocol based on ToF-like ranging techniques}
\label{sec:ToF}

In this chapter, we first discuss how faking sensors can corrupt the two SToF and DAT ranging techniques:
\begin{enumerate}
\item In case the \emph{SToF} ranging technique is used by Sensor $u$, $u$ first transmits a message attaching the time of transmission $t_s$ into the message.
Sensor $v$, which receives the message from sensor $u$ at time $t_r$, estimates the distance based on delay $t=t_r - t_s$ and radio signal propagation speed $s_r$, $\hat{d}(v,u)=s_rt$.
So, it is possible that a faking sensor can prevent sensor $v$ from computing the real distance by faking the transmission time $t_s$.
\item In case the \emph{DAT} ranging technique is used, Sensor $u$ transmits each message simultaneously with two signals (\emph{e.g.} RF and US signals). 
Sensor $v$ then records the difference of arrival time $t$ between RF signal and US signal.
This can be done using only a local clock at $v$. Thus no global time is required.
Then, Sensor $v$ computes distance $\hat{d}(v,u)$ based on $t$, propagation speed $s_r$ of RF signal and propagation speed $s_u$ of US signal.
In this case, a faking sensor may prevent a correct sensor $v$ from computing real distance by delaying one of the two simultaneous transmissions.

\end{enumerate}

Now we show that corrupting SToF and DAT ranging technique has the same affect on correct sensors.

\begin{lemma}
\label{lem:SToF}
If the ranging is evaluated with SToF technique and faking sensor $F$ shifts real transmission time then all correct sensors compute the real distance to sensor $F$ increased or decreased by the same length $b$.
\end{lemma}

\begin{proof}
Let us assume that faked sensor $F$ shifts its real transmission time by $t'$.
Then all the correct sensors will compute the distance
modified by $b=s_rt'$, where $s_r$ is the radio signal propagation speed.
\end{proof}

\begin{lemma}
\label{lem:DAT}
If the ranging is evaluated with DAT technique and faking sensor $F$ introduces shift $t'\neq 0$ between the RF and US transmissions, then all correct sensors compute the real distance to the sensor $F$ increased or decreased by the same length $b$.
\end{lemma}

\begin{proof}
Since the faking sensor shifts the two transmissions by time $t'$ then the difference in
arrivals time of the signals will be $t+t'$ where $t$ is original difference for $t'=0$.
Each correct sensor will compute $\hat{d}$ based on the following equation.
\[
t+t' = \frac{\hat{d}}{s_r} - \frac{\hat{d}}{s_u}
\]

Thus the real distance will be modified by
\[
b= \frac{t'}{1/s_r - 1/s_u}
\]
in all correct sensors.
\end{proof}

Since the corruption on SToF and DAT has the same result we can formulate the following theorem for both ranging techniques. 

\begin{theorem}
\label{Thm_SToF_DAT}
If the distance evaluation is done with SToF or DAT techniques and no six sensors are located on the same hyperbola and no three sensors are co-linear, then at least one of six correct sensors detects inconsistency in faked transmission.
\end{theorem}

\begin{proof}

Let us assume that faking sensor $F$ enlarges its distance against the correct sensors
by $b$. The case when sensor reduces its distance is symmetric.
By lemma \ref{lem:SToF} and  \ref{lem:DAT} there are at most two faked locations $F'$ and $F''$ for faking sensor $F$,
which guarantee consistency against sensors $P_1$ and $P_2$ (see figure \ref{fig:figToF}).
Let us assume that sensor $F$ decides for faked location $F'$.

\begin{figure}[hbp]
\begin{center}
\epsfig{figure=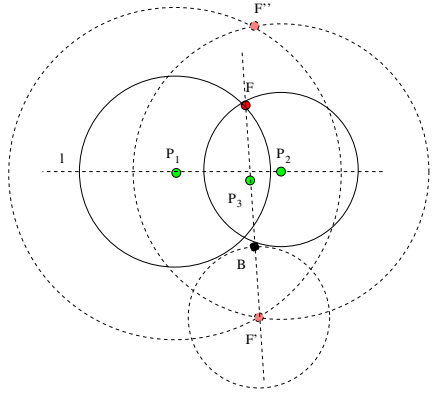 , width=0.4\textwidth}
\caption{\label{fig:figToF} Figure shows that sensor $F$ can change its position to $F'$
and consistently lie against sensor $P_3$ which is located in the middle of segment $FB$.
Length of segment $F'B$ is $b$. }
\end{center}
\end{figure}

Now we will find the set of correct sensors, which will not detect the inconsistency.
We consider two cases:
\begin{enumerate}

\item
%(see figure \ref{fig:figToF}).
The first case is when distance $c$ between $F'$ and $F$ is strictly larger than $b$ (see figure \ref{fig:figToF1}).
Each correct sensors $P$, which cannot detect inconsistency on distance to $F$, has to meet $d(P,F)=\hat{d}(P,F)$.
The condition $d(P,F)=\hat{d}(P,F)$ can be transformed into the distances on the plane $|F'P|= |FP| + b$.
Based on this condition we can came up with system of equations for sensors in $S=\{P: d(P,F)=\hat{d}(P,F) \}$.

\begin{figure}[htbp]
\begin{center}
\epsfig{figure=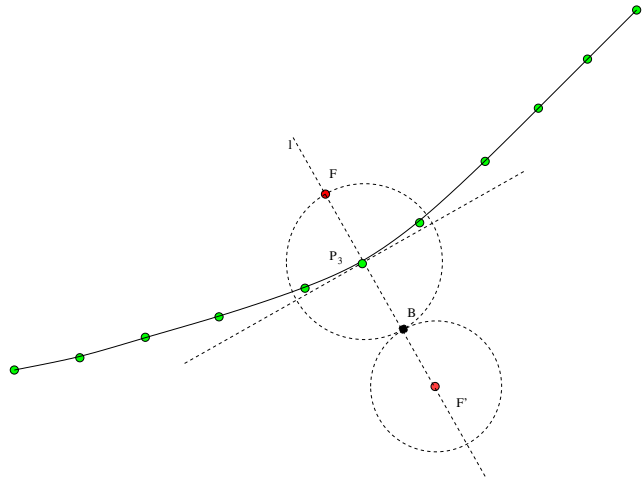 , width=0.6\textwidth}
\caption{\label{fig:figToF1}
We assume that $|FF'| >b$.
Figure shows set $S$ of correct sensors located on the hyperbola, which cannot detect inconsistency.
That is for each correct sensor $P$ located on the hyperbola the distance $|F'P|$ is equal to $|FP|+b$
}
\end{center}
\end{figure}

\begin{eqnarray}
x^2 + y^2 & = & z^2 \nonumber \\
x^2 + (y - c)^2 & = & (z+b)^2 \nonumber \\
\end{eqnarray}

Where $|FP|=z$, $x,y$ are the coordinates of correct sensor $P \in S$.
We assume that $F=(0,0)$ and $F'=(0,c)$.
Next we can find the equation of the hyperbola.

\begin{eqnarray}
x^2 + (y - c)^2 & = & (\sqrt{x^2 + y^2}+b)^2 \nonumber \\
x^2 + y^2 - 2yc + c^2 & = & x^2 + y^2 + 2b\sqrt{x^2 + y^2} + b^2 \nonumber \\
(-2yc + c^2 - b^2)^2 & = &  4b^2(x^2 + y^2)  \nonumber \\
4y^2c^2 - 4yc(c^2-b^2) + (c^2-b^2)^2  & = &  4b^2(x^2 + y^2)  \nonumber \\
4y^2c^2 - 4yc(c^2-b^2) + (c^2-b^2)^2  - 4b^2 y^2 & = & 4b^2x^2  \nonumber \\
4(c^2-b^2)y^2 - 4c(c^2-b^2)y + (c^2-b^2)^2 & = & 4b^2x^2  \nonumber \\
(c^2-b^2) ( 4y^2 - 4cy + c^2-b^2) & = & 4b^2x^2  \nonumber \\
(c^2-b^2) ( (2y-c)^2 - b^2 ) & = & 4b^2x^2  \nonumber \\
(c^2-b^2) (2y-c)^2 - b^2 (c^2-b^2)  & = & 4b^2x^2  \nonumber \\
(c^2-b^2) (2y-c)^2 - 4b^2x^2   & = & b^2 (c^2-b^2)  \nonumber \\
(c^2-b^2) (2y-c)^2 - 4b^2x^2 & = & b^2 (c^2-b^2)  \nonumber \\
\frac{(2y-c)^2}{b^2} - \frac{4x^2}{c^2-b^2}  & = &  1
\end{eqnarray}

The five sensors uniquely determine the hyperbola.
Thus the sixth sensor, which is not located on the hyperbola by our assumption,
will detect inconsistency.
 
\item
The second case is when distance $c$ between $F'$ and $F$ is at most $b$ (see figure \ref{fig:figToF2}). 
We will show that $P_1$ or $P_2$ will have to detect inconsistency.
The distance measured using coordinates by $P_i$ for $i=1,2$ has to be exactly
$|FP_i|+b$ to prevent sensor $P_i$ from detecting inconsistency.
By triangle inequality we have $|F'F| + |FP_i| \geq |F'P_i|$ for $i=1,2$.
Thus the distance $|F'P_i|$ measured by $P_i$ with a ranging technique is at most $|FP_i|+b$.
Sensor $P_i$ for $i=1,2$ will measure required distance when sensors $F'$, $F$ and $P_i$ are co-linear.
This will happen for at most one sensor.
This is because we assume that no three sensors are co-linear.
\begin{figure}[htbp]
\begin{center}
\epsfig{figure=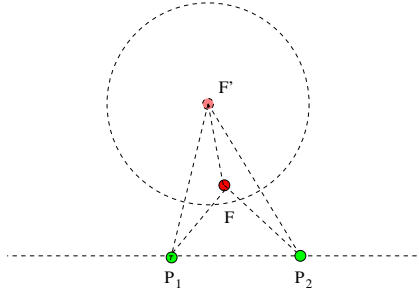 , width=0.4\textwidth}
\caption{\label{fig:figToF2}
We assume $FF' \leq b$
Figure shows that faking sensor $F$ cannot change its position to $F'$ consistently against sensors $P_1$ and $P_2$.
That is $F'P_1 < |FP_1| + b$ or $F'P_2 < |FP_2| + b$ allowing sensor $P_1$ or $P_2$ to detect inconsistency.}
\end{center}
\end{figure}
\end{enumerate}
\end{proof}

Theorem \ref{Thm_SToF_DAT} allows us to modify the protocol \textsc{FindMap} so that it works in the model in which faking sensors can corrupt the SToF or DAT ranging technique.

\begin{corollary}
\label{Cor_SToF_DAT_FIndMap}
The protocol \textsc{FindMap($\lceil \frac{n}{2} \rceil -2$)} identifies all faking sensors, in the model where faking sensors can corrupt SToF or DAT ranging techniques,
provided $n-f-5 > f$ and no six sensors are located on the same hyperbola and no three sensors are co-linear.
\end{corollary}

\begin{proof}
Let us consider a faking sensor $F$.
Theorem \ref{Thm_SToF_DAT} guarantees that in each set of six correct
sensors there exists a sensor which detects inconsistency on distance to $F$.
Thus each faking sensor will be accused by at least correct $n-f-5$ sensors.
By inequality $n-f-5 > f$ the number of correct sensors that accuse $F$ is at least $\lceil \frac{n}{2} \rceil - 2$
and the number of faking sensors is at most $\lfloor \frac{n}{2} \rfloor - 3$.
Thus each faking sensor will be find faking and no correct sensor will be found faking.
\end{proof}

Theorem \ref{Thm_SToF_DAT} can be also applied in the protocol for the model of trusted sensors~\cite{CCS06}. 
We can use theorem \ref{Thm_SToF_DAT} to compute the deployment of the minimum number of hidden stations required to detect faking nodes.

\begin{corollary}
\label{hiddenBase_SToF_DAT_FIndMap}
If the six hidden stations are not located on the same hyperbola and no three stations are co-linear then one of the stations always detect a faking node.
\end{corollary}

Corollary \ref{hiddenBase_SToF_DAT_FIndMap} is true provided the attacker's transmission reaches all the hidden stations and attacker is not allowed to use directional antennas.
Since the verifiers are hidden to the faking node, the latter has very low chance to consistently fake its position even with directional antennas.

\chapter{Concluding Remarks}

We proposed a secure positioning deterministic protocol for WSN that performs in the most general NTS model.
Although the previous protocol of Hwang \emph{et al.} \cite{HHK07} is probabilistic (and thus, unlike ours, can not give \emph{certain} results), it is interesting to see if the certainty of the result comes with a price (with respect to the number of exchanged messages to solve the problem). 
In~\cite{HHK07}, each sensor announces one distance at a time in a round robin fashion (otherwise the faking node could hold its own announcement, collect all correct nodes informations, and send a consistent range claim), inducing $n(n-1)$ sent messages, an overall $O(n^2)$ message complexity. In our case, $n$ coordinate messages are sent in round one, and $n$ accusation messages are sent in round two, overall a $O(n)$ message complexity. However, from a information complexity point of view, the two approaches are equivalent, since the exchanged messages in our protocol can be $n$-sized (inducing $n^2$ information in both cases).

To conclude, we would like to mention two interesting open questions:
\begin{enumerate}

\item Our protocol makes some synchrony hypotheses to separate between rounds and filter faking nodes. It is worth investigating to determine the exact model assumptions that are necessary and sufficient to solve the same problem in the NTS model with respect to synchrony.

\item Our network model assumes that correct nodes are within range of every other node. Extending our result to WSN with fixed ranges for every node is a challenging task, especially since previous results on networks facing intermittent failures and attacks~\cite{DT,DDT,NT} are written for rather stronger models (\emph{i.e.} wired secure communications) than that of this paper. 

\end{enumerate}

\bibliographystyle{plain}
\bibliography{mybib}

\end{document}